\documentclass[onecolumn,superscriptaddress,nobibnotes,nofootinbib,floatfix,aps,prl,citeautoscript,reprint]{revtex4-1}
\usepackage{graphicx}
\usepackage{tabularx}
\usepackage{booktabs}
\usepackage{lmodern}
\usepackage{dcolumn}
\usepackage{bm}
\usepackage{float}
\usepackage{siunitx}
\usepackage[version=4]{mhchem}
\usepackage{mathptmx}
\usepackage[colorlinks=true,linkcolor=blue,citecolor=blue,urlcolor=blue]{hyperref}
\usepackage[caption=false]{subfig}
\usepackage[usenames,dvipsnames,svgnaes,table]{xcolor}
\definecolor{col1}{rgb}{0.0, 0.46, 0.8}
\definecolor{col2}{rgb}{0.9, 0.0, 0.30}

\begin{document}

\title{Prediction of superconducting iron--bismuth intermetallic compounds at high pressure}

\newcommand{\ub}{Department of Physics, Universit\"{a}t Basel,
Klingelbergstr. 82, 4056 Basel, Switzerland}
\newcommand{\nwchem}{Department of Chemistry, Northwestern University, Evanston, Illinois 60208, USA}
\newcommand{\nwmse}{Department of Materials Science and Engineering, Northwestern University, Evanston, Illinois 60208, USA}
\newcommand{\nwearth}{Department of Earth and Planetary Sciences, Northwestern University, Evanston, Illinois 60208, USA}
\newcommand{\argonne}{High-Pressure Collaborative Access Team, Argonne National Laboratory, Carnegie Institution of Washington, Argonne, Illinois 60439, USA}

\author{Maximilian Amsler}
\affiliation{\nwmse}

\author{S. Shahab Naghavi}
\affiliation{\nwmse}

\author{Chris Wolverton}
\affiliation{\nwmse}
\email{c-wolvteron@northwestern.edu}

\begin{abstract}

The synthesis of materials in high-pressure experiments has recently attracted increasing attention, especially since the discovery of record breaking superconducting temperatures in the sulfur-hydrogen and other hydrogen-rich systems. Commonly, the initial precursor in a high pressure experiment contains constituent elements that are known to form compounds at ambient conditions, however the discovery of high-pressure phases in systems immiscible under ambient conditions poses an additional materials design challenge. We performed an extensive multi component \textit{ab initio} structural search in the immiscible Fe--Bi system at high pressure and report on the surprising discovery of two stable compounds at pressures above $\approx36~$GPa, \ce{FeBi2} and \ce{FeBi3}. According to our predictions, \ce{FeBi2} is a metal at the border of magnetism with a conventional electron-phonon mediated superconducting transition temperature of $T_{\rm c}=1.3$~K at 40~GPa. In analogy to other iron-based materials, \ce{FeBi2} is possibly a non-conventional superconductor with a real $T_{\rm c}$ significantly exceeding the values obtained within Bardeen-Cooper-Schrieffer (BCS) theory.

\end{abstract}

\maketitle

\section{\label{sec:intro}Introduction}

Improved strategies to discover energy materials are called for to tackle the inevitable global environmental challenges due to limited fossil fuels and climate change. Recent advances in materials science have not only been aimed at exploring uncharted chemical space, but has also brought forward novel synthesis pathways to design materials at non-ambient conditions. In addition to composition and temperature, pressure constitutes an accessible degree of freedom to be sampled in the search for novel materials. Significant progress has been made in high-pressure techniques such that several hundred GPa can be meanwhile readily achieved in diamond anvil cells (DAC).

Often, materials design rules based on chemical intuition derived at ambient conditions cannot be directly applied at high pressure, where unexpected physical phenomena can lead to surprising discoveries in novel compositions, bonding and electronic structures. \textit{Ab initio} calculations have proven to provide crucial insight in understanding and predicting new phases at these conditions. The discovery of an ionic form of boron for example was first predicted from evolutionary structural search and later confirmed by experiments~\cite{oganov_ionic_2009}, and similarly the metal-insulator transition in elemental sodium was intially predicted from density functional theory (DFT) calculations~\cite{ma_transparent_2009}. Recently, a range of unexpected stoichiometries was found in the Na-Cl system at high pressure with compositions ranging from \ce{NaCl3} to \ce{Na3Cl}~\cite{zhang_unexpected_2013}, radically defeating chemical intuition for ionic materials.

Many high pressure studies, including the examples above, are commonly performed with precursors (i.e. crystals or molecules) containing constituent elements that are known to form some compound at ambient condition. This choice is well justified due to two reasons: first, it is easier and hence preferable to place a sample into a DAC which already exhibits the targeted interatomic bonds. Second, the risk of elemental decomposition can be expected to be lower if the constituent elements form stable compounds at some known condition. Studying alloy systems at high pressures with severe immiscibility at ambient pressure (i.e. not forming compounds over any range of composition and temperature) therefore poses a significant additional materials discovery challenge. In fact, bismuth is well known for its notorious solid-state immiscibility, which has precluded the formation of binaries with a wide range of elements~\cite{boa_ternary_2008}, leading to various high pressure attempts to synthesize novel bismuth containing intermetallics~\cite{Clarke2016,Matthias1961,Schwarz2013,Tence2014}. In particular, the ambient phase diagram of the Fe--Bi intermetallic system shows essentially no solubility of Fe in Bi (or vice versa)~\cite{boa_ternary_2008} and thus constitutes an excellent example of a system possibly containing unexpected high-pressure phases awaiting discovery.

Superconductivity has been the main focus of many recent theoretical and experimental high-pressure studies, with an increasing interest in hydrogen-rich materials since the discovery of record-breaking transition temperatures in the range of 100-200~K in sulfur- and phosphorus-hydrides~\cite{Ashcroft_PRL1968,Ashcroft_PRL2004,Cudazzo_PRL2008,mcmahon_high_2011,Szcz_superconducting_2009,tse_novel_2007,Chen_PNAS2008,Kim_PNAS2008,FengAsHoffman_Nature2008,Wang_PNAS2009,Yao_PNAS2010,gao_high-pressure_2010,Kim_PNAS2010,Li_PNAS2010,Zhou_PRB2012,Hooper_JPC-2014,Duan_SciRep2014,SH_PRB-Mazin-2015,Errea_anhaPRL2015,PRB_Duan2015,PRB_akashi_2015_HS,Maramatsu-Hemley_2015,flores-livas_high-pressure_2012,drozdov_conventional_2015,flores-livas_high_2016,drozdov_superconductivity_2015,flores-livas_superconductivity_2016}. Similarly, iron based superconductors have recently been intensely studied~\cite{stewart_superconductivity_2011,johnson_iron-based_2015} in so called 1111~\cite{kamihara_iron-based_2008,han_srfeasf_2008}, 122~\cite{christianson_unconventional_2008}, 111~\cite{tapp_lifeas:_2008}, and 11~\cite{hsu_superconductivity_2008} compounds. Ferro pnictides such as LiFeAs~\cite{wang_superconductivity_2008,tapp_lifeas:_2008,chu_synthesis_2009} and Sr$_{0.5}$Sm$_{0.5}$FeAsF~\cite{wu_superconductivity_2009} exhibit high transition temperatures at ambient condition of $T_{\rm c}=18$~K and $T_{\rm c}=56$~K, respectively, while other compounds such as NaFeAs~\cite{parker_structure_2009,chu_synthesis_2009,zhang_superconductivity_2009} and FeSe~\cite{hsu_superconductivity_2008,medvedev_electronic_2009} show a strong increase in $T_{\rm c}$ at high pressure (e.g. from 8 to 36.7~K in FeSe). The superconducting mechanism in all these iron-based compounds is unconventional and thus not based on electron-phonon coupling~\cite{Haule2008,mazin_unconventional_2008,Mazin2010}, instead the proximity to magnetism suggests that magnetic (spin) fluctuations play a key role in mediating superconductivity~\cite{Haule2008,mazin_unconventional_2008,Mazin2010,Dai2012}. Furthermore, many phosphide, arsenide and antimonide superconductors have been discovered, also reviving intense investigations in bismuth containing compounds. The intermetallic compound Ca$_{11}$Bi$_{10-x}$ was found to be superconducting with $T_{\rm c}=2.2$~K, and several other Ca--Bi binaries were predicted to have $T_{\rm c}$s in the range of $2.27-5.25$~K in high pressure phases~\cite{dong_rich_2015}. The nickel--bismuth binaries, \ce{NiBi}~\cite{Haegg1929} and \ce{NiBi3}~\cite{Glagoleva1954,Ruck2006}, are both superconductors with $T_\mathrm{c}$ values of \SI{4.25}{\kelvin}~\cite{Alekseevskii1952} and \SI{4.06}{\kelvin}~\cite{Alekseevskii1948,Herrmannsdorfer2011}, respectively. Similarly, the \ce{CoBi3} high pressure compound is a superconductor with $T_\mathrm{c} = \SI{0.48}{\kelvin}$~\cite{Matthias1961,Schwarz2013,Tence2014}, as well as the copper--bismuth binary \ce{Cu11Bi7} which forms at high-pressure with a $T_\mathrm{c}$ of $\SI{1.36}{\kelvin}$~\cite{Clarke2016}.

Here we report on the prediction of two stable high-pressure compounds, \ce{FeBi2} and \ce{FeBi3}, in the completely immiscible Fe--Bi system by performing an extensive multi-component \textit{ab initio} structural search. The Fe--Bi system not only shows no stable compounds in its ambient-pressure phase diagram~\cite{boa_ternary_2008}, but there is virtually no solubility of either solid-state element in the other. Thus, the prediction of stable compounds in this system is particularly surprising. In contrast to \ce{FeSb2} and \ce{FeAs2}, which are both semiconductors with promising thermoelectric properties, \ce{FeBi2} is metallic in a wide pressure range. The ferromagnetic and antiferromagnetic order in \ce{FeBi2} is suppressed by pressure, leading to a superconducting behavior with a conventional $T_{\rm c}$ of 1.3~K in the non magnetic state at 40~GPa. Due to its proximity to magnetism, \ce{FeBi2} is possibly a new member in the family of unconventional iron-pnictide  superconductors~\cite{Mazin2010}.

\section{Method}

Density functional theory (DFT) calculations were carried out to predict the composition, structure, and properties of novel binary Fe-Bi compounds. The Minima Hopping structure prediction method (MHM) as implemented in the \texttt{Minhocao} package~\cite{goedecker2004,amsler2010} was employed to perform a multi-component search for stable phases at high pressure. The MHM implements a reliable algorithm to identify the ground state structure of any compound by efficiently sampling low lying phases on the enthalpy landscape, based solely on the information of the chemical composition~\cite{amsler_crystal_2012,flores-livas_high-pressure_2012,amsler_novel_2012}. Consecutive short molecular dynamics escape steps are performed to overcome enthalpy barriers followed by local geometry optimizations, while exploiting the Bell-Evans-Polanyi principle in order to accelerate the search~\cite{roy_bell-evans-polanyi_2008,sicher_efficient_2011}.

The energies, forces and stresses were evaluated from DFT calculations within the projector augmented wave~(PAW) formalism~\cite{PAW-Blochl-1994} as implemented in the \texttt{VASP}~\cite{VASP-Kresse-1995,VASP-Kresse-1996,VASP-Kresse-1999} code together with the Perdew-Burke-Ernzerhof (PBE) approximation~\cite{Perdew-PBE-1996} to the exchange correlation potential. A plane-wave cutoff energy of 400~eV was used in conjunction with a sufficiently dense k-point mesh to ensure a convergence of the total energy to within 1~meV/atom. Geometries were fully relaxed with a tight convergence criterion of less than 4~meV/\AA~for the maximal force components.

The magnetic properties for the estimation of the Stoner parameter were evaluated with the full potential linearized augmented plane wave (FLAPW) method as implemented in the \texttt{WIEN2k} code~\cite{blaha_full-potential_1990}. The number of plane waves was restricted by $R_{MT}k_{max}=9$. All self-consistent calculations were performed with 6000 k-points in the irreducible wedge of the Brillouin zone, based on a mesh of $18 \times 18 \times 18$ k-points. The convergence criteria were set to $10^{-5}$~Ry for the energies and simultaneously to $10^{-3}$~$e ̄ $ for charges.

Superconducting properties were computed with the \texttt{Quantum Espresso} package~\cite{espresso} together with ultra-soft pseudopotentials and a plane-wave cutoff energy of 60~Ry. The phonon-mediated superconducting temperature was estimated using the Allan-Dynes modified McMillan's approximation of the Eliashberg equation~\cite{Allen_1975} according to
\begin{equation}\label{eq:mcmillan}
  T_\text{c}=\frac{\omega_\text{log}}{1.2}\exp\left[-\frac{1.04(1+\lambda)}{\lambda-\mu^{*}(1+0.62\lambda)}\right]
\end{equation}
where $\lambda$ is the overall electron-phonon coupling strength computed from the frequency dependent Eliashberg spectral function $\alpha^2F(\omega)$, $\mu^{*}$ is the Coulomb pseudopotential, and $\omega_\text{log}$ is the logarithmic avarage phonon frequency. A $8\times 8\times 8$ $q$-mesh was used together with a denser $24\times 24\times 24$ $k$-mesh, resulting in well converged values of the superconducting transition temperature $T_{\rm c}$. A typical Coulomb pseudopotential of $\mu^*=0.13$ was employed, a value which was shown to give $T_{\rm c}$'s in excellent agreement with experimental results for other bismuth superconductors~\cite{Clarke2016}.

\section{\label{sec:res}Results and Discussion}

We employed the MHM within the DFT framework to fully assess the stability of high-pressure phases of the Fe-Bi system. A pre-screening of only few compositions showed that Fe-rich compositions were overall less stable, such that the Bi-rich region was more densely sampled. Overall, structural searches were conducted in the compositional space of Fe$_x$Bi$_{1-x}$ for $x=(0.2$, $0.\overline{2}$, 0.25, 0.3, $0.\overline{3}$, 0.375, 0.4, $0.\overline{428571}$, $0.\overline{4}$, 0.5, 0.6, $0.\overline{6}$, 0.75) with up to 4 formula units per cell at 50 GPa, scanning several thousand different structures. The initial seeds were randomly generated or taken from already known bismuth intermetallics whenever available in structural databases. A range of the lowest energy structures at each compositions were subsequently relaxed with refined parameters at pressures between 0 and 100~GPa to obtain the complete pressure-composition phase diagram.

\begin{figure}[tb]
\centering
\subfloat[Convex hull of stability]{%
  \includegraphics[width=\columnwidth,angle=0]{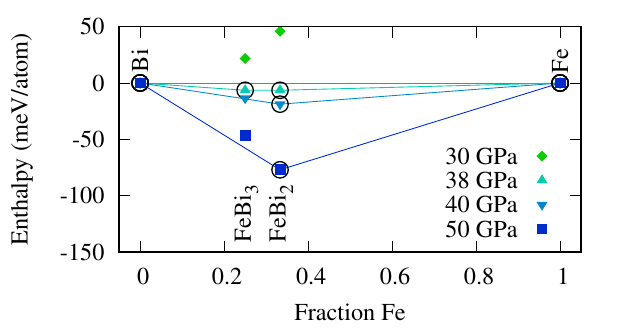}
}\\
% \begin{subfigure}[b]{1\columnwidth}\includegraphics[width=\columnwidth,angle=0]{Hulls-eps-converted-to.pdf}
% \caption{Convex hull of stability}
% \end{subfigure}
\subfloat[Pressure range of stability]{%
  \includegraphics[width=\columnwidth,angle=0]{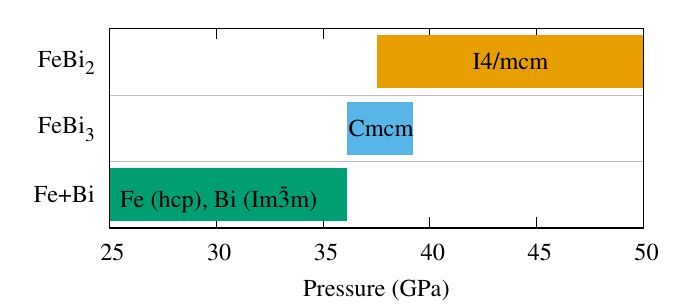}%
}
% \begin{subfigure}[b]{1\columnwidth}\includegraphics[width=\columnwidth,angle=0]{Stability-eps-converted-to.pdf}	
% \caption{Pressure range of stability}
% \end{subfigure}

\caption{\label{fig:stab} Panel (a) shows the formation enthalpies and the convex hull of stability as a function of Fe content for various pressures. The circles denote a compound that lies on the convex hull of stability. Panel (b) indicates the pressure range in  which FeBi$_2$ and FeBi$_3$ are thermodynamically  stable:  the  bottom  line  shows  the  range  in  which decomposition into elemental Fe and Bi is favored.}
\end{figure}

No thermodynamically stable compound was found up to around 36~GPa, at which point two binary phases, FeBi$_2$ and FeBi$_3$, exhibit negative formation enthalpies. The pressure range for which the compounds are thermodynamically stable are shown in Figure~\ref{fig:stab} together with the evolution of the convex hull of stability as a function of pressure. The range of stability for the FeBi$_3$ phase is rather narrow, merely between 36.1 and 39.2~GPa, whereas FeBi$_2$ remains thermodynamically stable from 37.5 up to at least 100~GPa. In fact, the magnitude by which the formation enthalpy of FeBi$_3$ is negative is very small, as shown in panel (a) of Figure~\ref{fig:stab}, such that the driving force for forming this phase is weak and it might be hard to experimentally synthesize it from elements. The FeBi$_2$ phase was predicted to crystallize in the \ce{Al2Cu} structure with space group \textit{I}4/\textit{mcm}. The lattice parameters at 40~GPa are $a=6.12$~\AA{\,} and $c=5.46$~\AA, respectively, with Fe and Bi at the Wyckoff positions $4a (0,0,0.250)$ and $8h (0.333,0.833,0)$, respectively. The FeBi$_3$ phase crystallizes in the \ce{PuBr3} structure~\cite{zachariasen_crystal_1948} with space group \textit{Cmcm} and lattice parameters $a=3.15$~\AA, $b=11.39$~\AA, and $c=7.93$~\AA, with Fe at the Wyckoff positions $4c (0,0.733,0.250)$, and two Bi at $8f (0,0.359,0.440)$ and $4c (0,0.059,0.250)$.

\begin{figure}[tb]
\includegraphics[width=0.52\columnwidth]{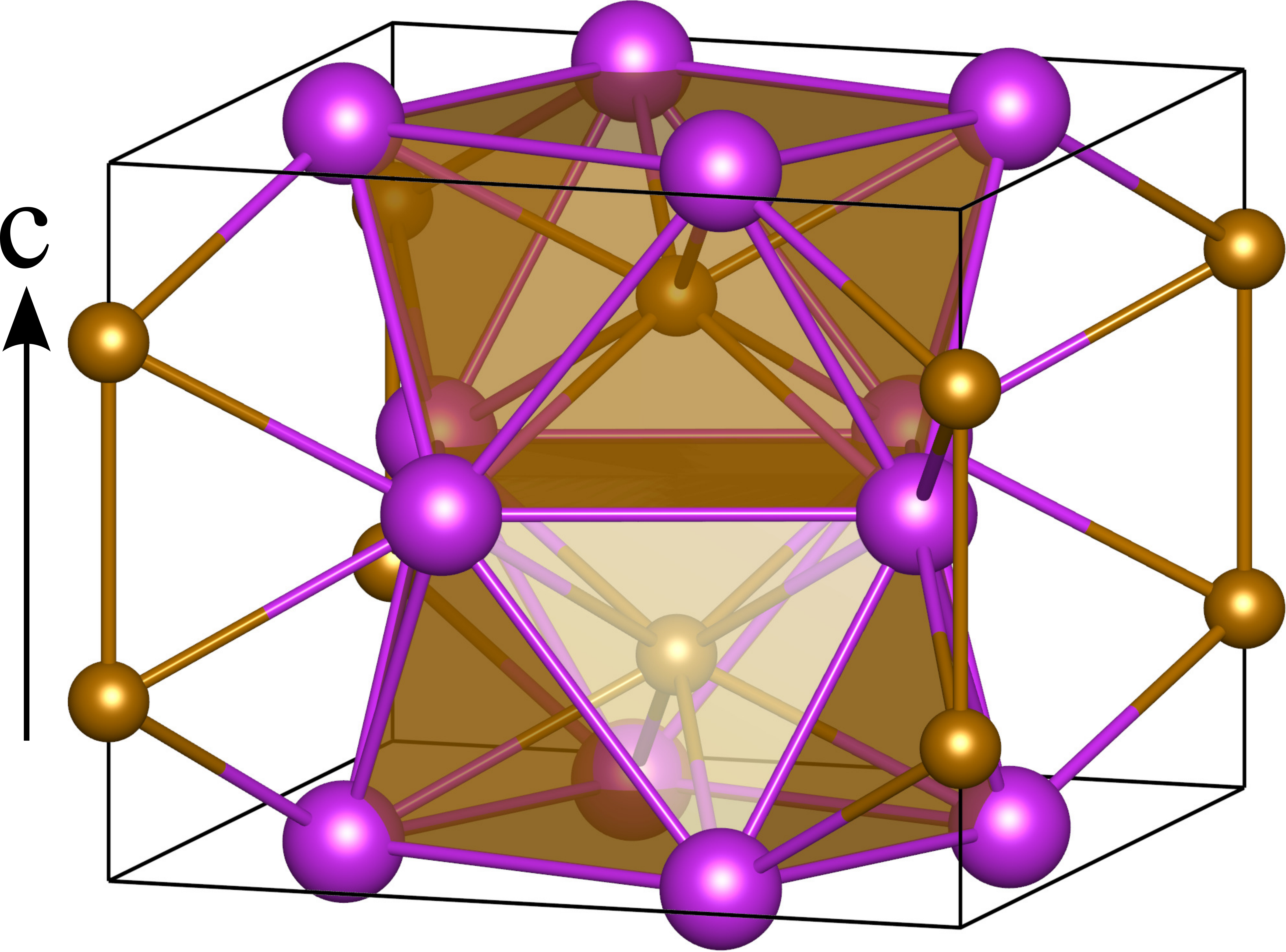}
\hfill
\includegraphics[width=0.43\columnwidth]{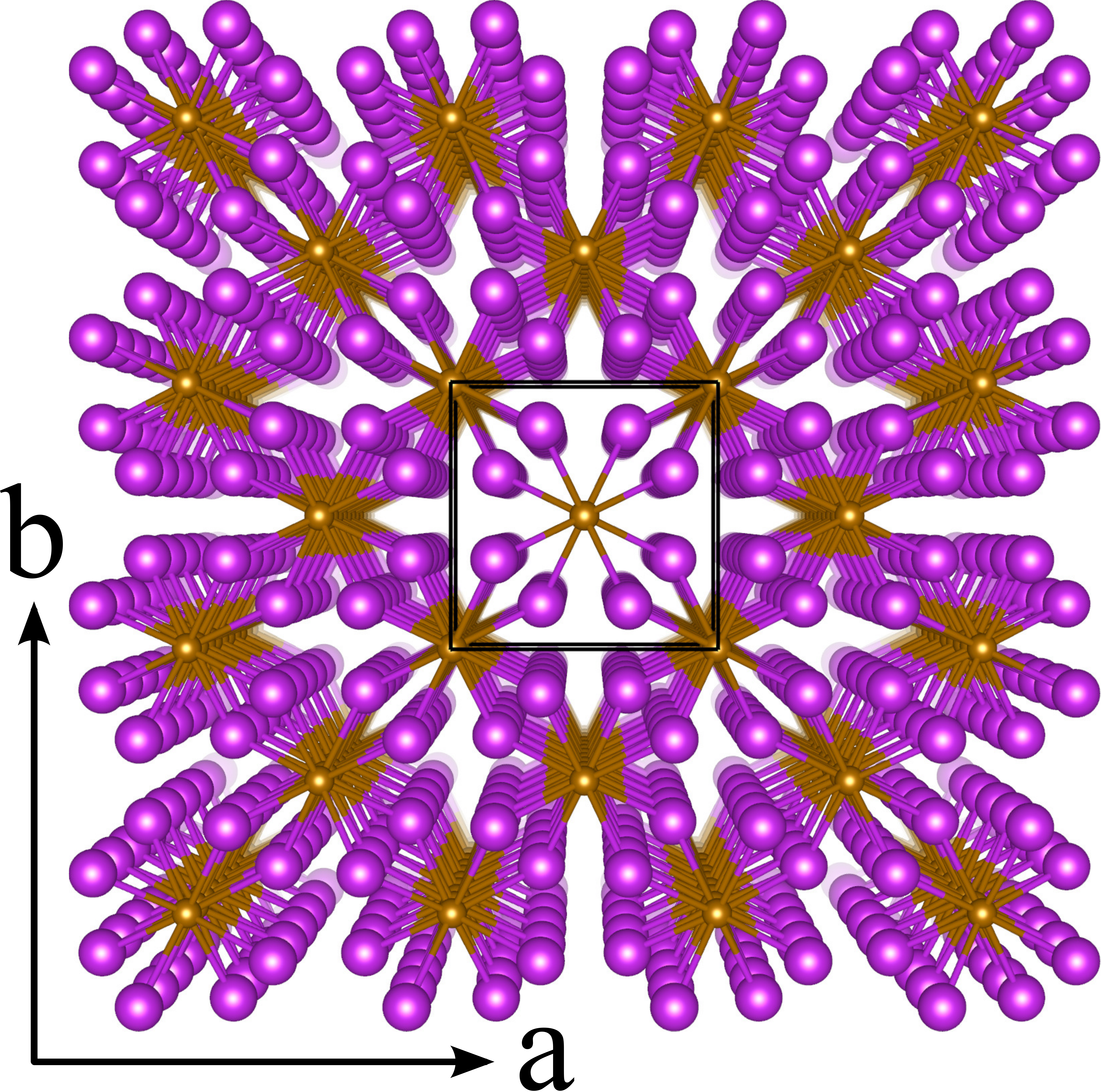}
\caption{\label{fig:MS_structure}Crystallographic   structure    of   \ce{FeBi2} optimized under ambient pressures. Left: view of two face-sharing \ce{\{FeBi8\}} square antiprisms  stacking in  the \textit{c}-direction. Right: view  down the \textit{c}-axis  showing the  edge-sharing linkages  formed between  the stacked columns.}
\end{figure}

During the structural search the well known marcasite phase of \ce{FeBi2} with space group \textit{Pnnm} was also recovered, which is the ground state structure of many iron-pnictides systems such as FeSb$_2$~\cite{hulliger_marcasite-type_1963}. In fact, the ICSD contains only two early transition metal--antimonides, \ce{TiSb2}~\cite{Nowotny1951} and \ce{VSb2}~\cite{Nowotny1951}, which crystallize directly in the \ce{Al2Cu} structure  (\textit{I}4/\textit{mcm}), but 9 further $3d$ transition metal pnictides \ce{MPn2} which attain the marcasite structure under ambient condition, namely \ce{CrSb2}~\cite{Haraldsen1949}, \ce{FeP2}~\cite{Meisel1934}, \ce{FeAs2}~\cite{Buerger1932}, \ce{FeSb2}~\cite{Hagg1928}, \ce{CoAs2}~\cite{Roseboom1963}, \ce{CoSb2}~\cite{Furst1938}, \ce{NiAs2}~\cite{Kaiman1947}, \ce{NiSb2}~\cite{Rosenqvist1953}, and \ce{CuAs2}~\cite{Peacock1940}.  Two of above \textit{Pnnm} compounds, \ce{CrSb2} and \ce{FeSb2}, have been shown experimentally to undergo a pressure-induced phase transition into the \ce{Al2Cu} structure at around \SI{5.5}{\giga\pascal}~\cite{Takizawa1999} and \SI{14.3}{\giga\pascal}~\cite{Poffo2012}, respectively. While these structural transitions have also been confirmed computationally~\cite{kuhn_electronic_2013}, the transition pressure in \ce{FeSb2} is slightly overestimated (38~GPa)~\cite{wu_pressure-induced_2009}. In analogy to these two compounds, the formation enthalpy of the marcasite structure in \ce{FeBi2} becomes lower than the \ce{Al2Cu} phase at pressures below 11~GPa, however it remains positive at all pressures and this phase is therefore thermodynamically unstable at any condition. Similarly, for the \ce{FeBi3} compound the \ce{RhBi3}-type structure with space group $Pnma$, which has also been reported in \ce{NiBi3}~\cite{fjellvag_structural_1987}, is thermodynamically favored with respect to the \ce{PuBr3} phase at pressures below 32~GPa but retains a positive formation enthalpy.

%Although \ce{Fe-Bi} bonds are rare in solid-state literature and non-existent in intermetallic compounds, which is perhaps unsurprising given the significant immiscibility issues in the iron--bismuth system. The shortest \ce{Bi-Fe} interatomic distance reported so far in a compound is \SI{2.99(1)}{\angstrom}, which is found in the ambient pressure crystal structure of bismuth ferrite, \ce{BiFeO3}~\cite{Sharma2014}. This compares well with the value obtained here for \ce{FeBi2} (\SI{2.99}{\angstrom}).

Since the composition with the largest range of stability is \ce{FeBi2}, we will henceforth focus on this compound in the \ce{Al2Cu} structure. Although there are many different interpretations of this structure~\cite{armbruster_bindungsmodelle_2005}, K. Schubert describes it as a stacking of square antiprisms along the \textit{c}-direction of the conventional cell~\cite{schubert_kristallstrukturen_1964}. Each antiprism consists of an iron atom which is surrounded by eight symmetrically equivalent bismuth atoms at identical interatomic distances of \SI{2.99}{\angstrom} at 0~GPa (see left panel in Figure~\ref{fig:MS_structure}). These antiprisms are stacked on top of each other by sharing their square faces, forming columns along the \textit{c}-direction and leading to \ce{Fe-Fe} distances of \SI{2.85}{\angstrom} at 0~GPa. These columns themselves are arranged in a square lattice within the \textit{ab}-plane (Figure~\ref{fig:MS_structure}, right panel) by sharing the edges of the antiprisms. The three unique \ce{Bi-Bi} bonds in \ce{FeBi2} form the edges of the square faces (\SI{3.72}{\angstrom}), the sides of the triangular faces (\SI{3.66}{\angstrom}), and the inter-column bonds in the \textit{ab}-plane (\SI{3.26}{\angstrom}).

We carried out a detailed theoretical investigation of the \ce{FeBi2} phase with respect to the chemical bonding, magnetic and superconducting properties  based on \textit{ab initio} calculations. Unusual magnetism is prevalent in several iron containing intermetallics with the \ce{Al2Cu} structure: \ce{FeGe2} was for example initially reported to be antiferromagnetic and ferromagnetic above and below 190~K~\cite{yasukochi_magnetic_1961}, respectively, but later studies could not reproduce the ferromagnetic state, reporting temperature dependent transitions from the paramagnetic state to spin spiral and collinear antiferromagnetism~(see Refs.~\onlinecite{corliss_magnetic_1985,jeong_electronic_2007} and references therein). Similarly, \ce{FeSn2} was reported to exhibit temperature dependent collinear and non-collinear antiferromagnetism~\cite{venturini_magnetic_1985}. Iron pnictides were found to exhibit temperature or pressure induced transitions from semiconductor to metal, accompanied with strong magnetic fluctuations~\cite{petrovic_anisotropy_2003,lukoyanov_semiconductorferromagnetic-metal_2006,perucchi_optical_2006}. Although these compounds crystallize in the marcasite phase, theoretical result predicts that \ce{FeP2}, \ce{FeAs2} and \ce{FeSb2} transform into the \ce{Al2Cu} structure at pressures of above 108, 92 and 38~GPa, respectively~\cite{wu_pressure-induced_2009}, and experimental observations report that the phase transition in \ce{FeSb2} indeed occurs at 14.3~GPa ~\cite{Poffo2012}.

\begin{figure}[tb]
\includegraphics[width=0.9\columnwidth]{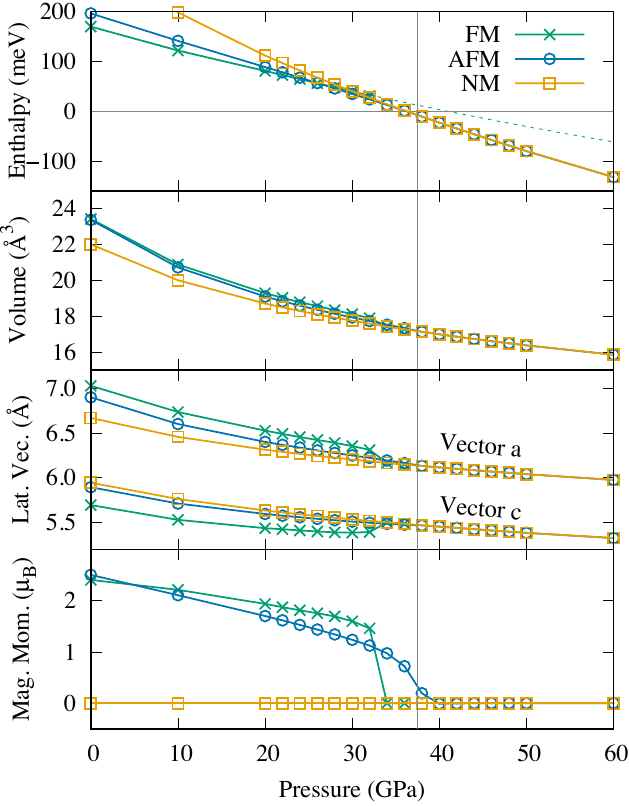}
\caption{\label{fig:Magvol} 
The top panel shows the formation enthalpy of \ce{FeBi2} in the FM, AFM and NM configuration (in meV/atom). The dashed line serves as a guide to the eye and was obtained from a quadratic fit to the enthalpy of the FM state between 0 and 30~GPa, before the magnetic collapse. The second panel shows how the volume per atom evolves as a function of pressure for the three spin configurations, whereas the third panel illustrates how the lattice vectors change. The magnetic moment per Fe as a function of pressure is shown in the bottom panel. The vertical gray line denotes the transition pressure above which \ce{FeBi2} becomes thermodynamically stable}
\end{figure}

To account for the various reported magnetic properties, we considered the closed shell non magnetic (NM) and two collinear magnetic states in this work: the ferromagnetic (FM) and one anti-ferromagnetic (AFM) configuration, where neighboring Fe atoms carry alternating spins as illustrated in Figure~2 of Ref.~\onlinecite{yasukochi_magnetic_1961}. Figure~\ref{fig:Magvol} shows how various materials properties vary as a function of pressure for the three different magnetic states. The thermodynamically most stable state at ambient condition is FM although it has a positive formation enthalpy, as illustrated in the top panel. Upon compression, the formation enthalpies of all three magnetic states gradually decreases, until at around 26~GPa AFM becomes the energetically most favorable state. Similarly, the AFM configuration competes with the NM state until at above 38~GPa when the NM state becomes the most stable. The magnetic moments as a function of pressure is shown in the bottom panel of Figure~\ref{fig:Magvol}. At ambient pressure, the magnetic moment in the FM configuration is 2.41~$\mu_B$ per Fe, whereas it is 2.51~$\mu_B$ per Fe for the AFM configuration. In both cases, the absolute value of the magnetic moment decreases monotonically as a function of pressure. At a critical pressure of 32~GPa for FM and 40~GPa for AFM, respectively, the magnetic spin polarization collapses, leading to the NM configuration. In contrast to AFM where the magnetic moment decreases smoothly, the spin collapse occurs discontinuously for the FM configuration, accompanied with a sudden decrease in the atomic volume and change in the cell parameters as illustrated in the two middle panels of Figure~\ref{fig:Magvol}.

In fact, this reduction in volume plays a crucial role for the stability of \ce{FeBi2}. The dashed line in the top panel of Figure~\ref{fig:Magvol} was obtained through a quadratic fit within the range of 0 to 30~GPa of the FM state and shows how the formation enthalpy would evolve if the magnetic collapse didn't occur. The formation enthalpy would stay positive until slightly above 40~GPa, and retains a slope with a magnitude much lower compared to the NM configuration. Consequently, the \ce{FeBi3} compound would compete with \ce{FeBi2} up to a much higher pressure than shown in Figure~\ref{fig:stab}, leading to a larger stability range of \ce{FeBi3} (and a smaller stability range of \ce{FeBi2}). Therefore, the reduction in volume due to the magnetic collapse is the main driving force that stabilizes FeBi$_2$ since the pressure term $pV$ in the enthalpy, $H=E+pV$, increasingly dominates the formation enthalpy at high pressure. Its decrease is essentially responsible for the thermodynamic stability of FeBi$_2$.

\begin{figure}[t]
\setlength{\unitlength}{1cm}
\subfloat[0~GPa: FM]{%
\includegraphics[height=5.1cm,angle=0]{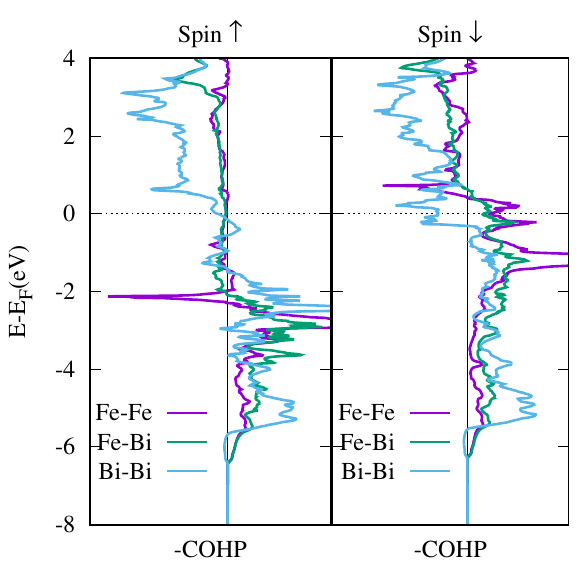}
}
\subfloat[0~GPa: NM]{%
 \includegraphics[height=5.1cm,angle=0]{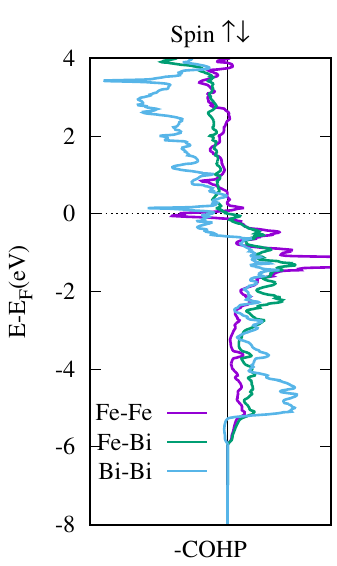}
}\\
\subfloat[30~GPa: FM]{%
\includegraphics[height=5.1cm,angle=0]{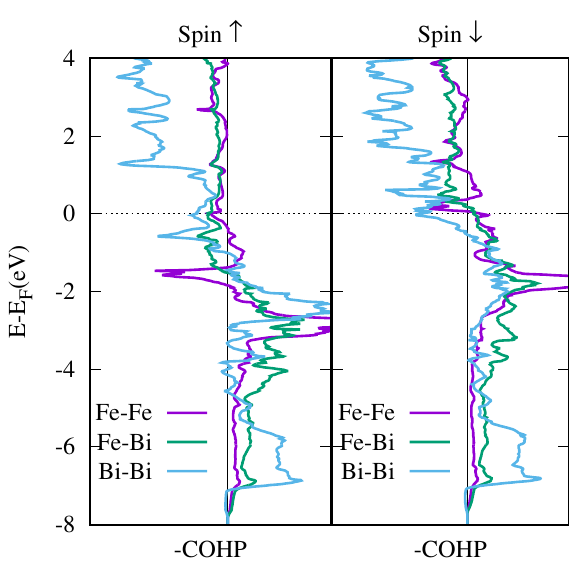}
}
\subfloat[40~GPa: NM]{%
\includegraphics[height=5.1cm,angle=0]{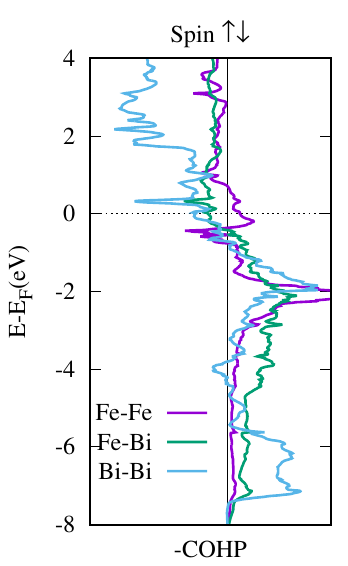}
}
% \subcaptionbox{0~GPa: FM}{\includegraphics[height=5.1cm,angle=0]{000GPa_cohp-eps-converted-to.pdf}}
% \subcaptionbox{0~GPa: NM}{
% \includegraphics[height=5.1cm,angle=0]{000GPa_NonMag_cohp-eps-converted-to.pdf}}	
% \subcaptionbox{30~GPa: FM}{\includegraphics[height=5.1cm,angle=0]{030GPa_cohp-eps-converted-to.pdf}}
% \subcaptionbox{40~GPa: NM}{\includegraphics[height=5.1cm,angle=0]{040GPa_NonMag_cohp-eps-converted-to.pdf}}
% \caption{The COHP for the Fe-Fe, Fe-Bi and Bi-Bi interactions at various pressures and spin configurations of \ce{FeBi2}. Panels (a) and (b) show the COHP for the spin polarized FM and closed shell NM configuration at 0~GPa, respectively. Panel (c) shows the FM configuration at 30~GPa, whereas panel (d) shows the NM configuration at 40~GPa.}
\label{fig:cohp}
\end{figure}

Based on above observations, the collapse of the magnetic state is evidently accompanied by a change in the bonding properties of FeBi$_2$. To analyze the interatomic bonding the crystal orbital Hamilton overlap population (COHP) was computed using the ~\texttt{LOBSTER} package~\cite{deringer_crystal_2011,dronskowski_crystal_1993,maintz_analytic_2013}. The bonding and antibonding states for the shortest Fe-Fe, Fe-Bi and the Bi-Bi bonds are plotted in Figure~\ref{fig:cohp}. For the NM configuration at 0~GPa shown in panel (b), where the two spin channels are equal (closed shell), the Fermi level falls in the antibonding region of both the Fe-Fe and Bi-Bi interactions, leading to an electronic instability. This unfavorable bonding is relieved in the spin polarized FM configuration shown in (a), where the antibonding states at the Fermi level for the $\uparrow$-spin channel are completely removed. When the structure is compressed, the Fermi level is gradually pushed into the antibonding region of both spin channels as shown in Figure~\ref{fig:cohp}~(c) for 30~GPa. At this point, the NM configuration becomes favorable and the system is driven towards a closed shell system where the Fermi level does not lie in the Fe-Fe antibonding states, as shown in Figure~\ref{fig:cohp}~(d) at 40~GPa.

\begin{figure}[b]
\setlength{\unitlength}{1cm}
\subfloat[30~GPa]{%
\includegraphics[width=0.49\columnwidth,angle=0]{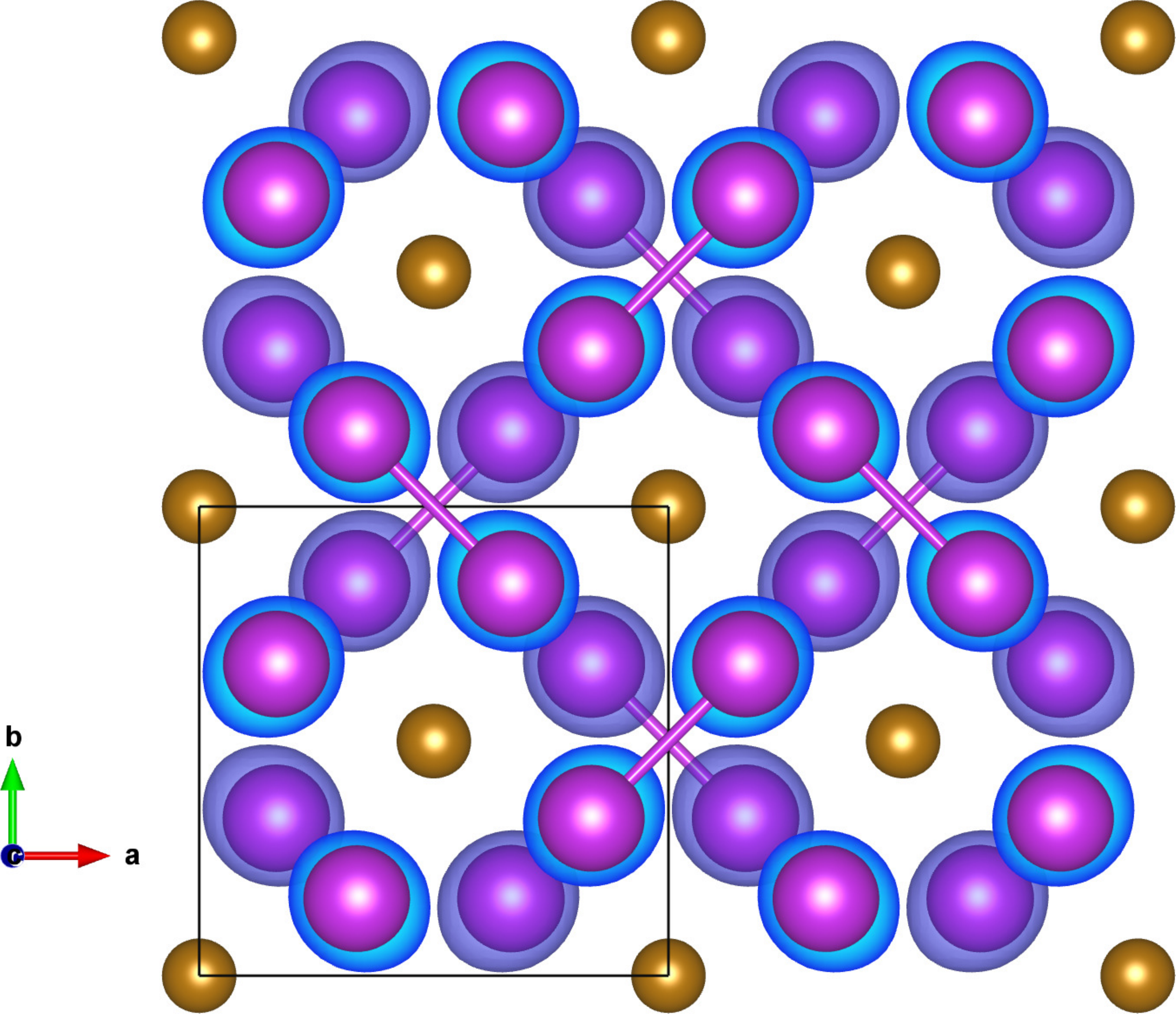}
}
\subfloat[40~GPa]{%
\includegraphics[width=0.49\columnwidth,angle=0]{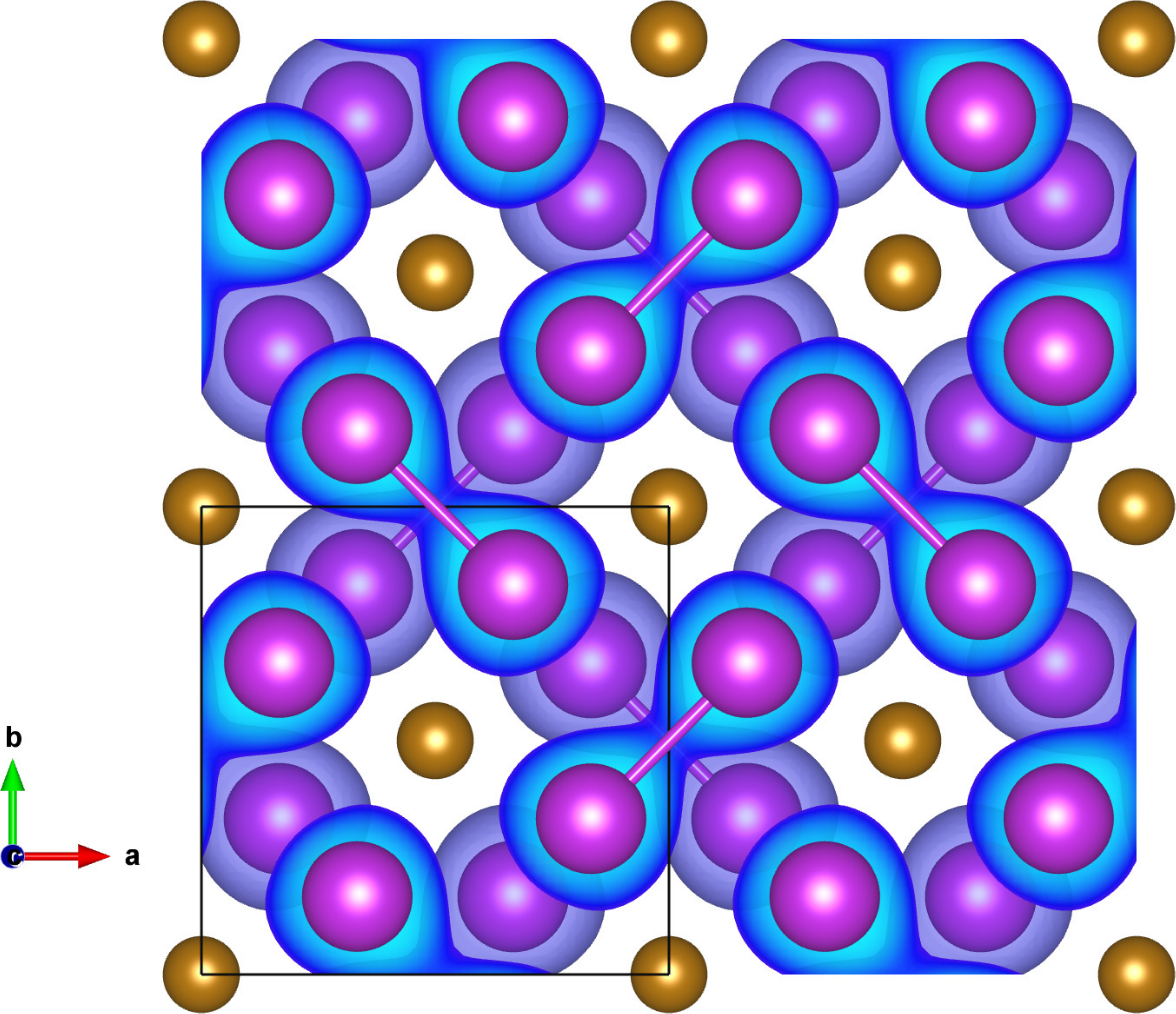}
}
% \subcaptionbox{30~GPa}{\includegraphics[width=0.49\columnwidth,angle=0]{ELFCAR_trim1-eps-converted-to.pdf}}
% \subcaptionbox{40~GPa}{\includegraphics[width=0.49\columnwidth,angle=0]{ELFCAR_trim2-eps-converted-to.pdf}}
\caption{\label{fig:elf}Electron localization function (ELF) at a value of 0.6 at 30 and 40~GPa, where panel (a) shows the $\uparrow$-spin channel of the FM configuration and panel (b) shows the NM configuration. The gold (small) spheres denote Fe atoms, while the purple (large) spheres denote Bi atoms. The section in the x-y plane is shown to illustrate the gradient of the ELF.}
\end{figure}

This change in the bonding properties can also be observed when analyzing the electron localization function (ELF). Figure~\ref{fig:elf} shows the ELF within the Bi layers of the FM and NM configuration at 30 and 40~GPa, respectively. The electrons, which are initially localized on the individual atoms (see panel (a)), are transferred to the Bi layers to form Bi-Bi dumbbells with strongly covalent character and electrons localized between the Bi atoms. Simultaneously, the Fe-Fe bond is weakened as evident by the increasing lattice constant in the $c$-direction (Figure~\ref{fig:Magvol}, panel 3). This behavior is in good agreement with the COHP shown in  Figure~\ref{fig:cohp}, where the Bi--Bi antibonding states at the Fermi level are reduced upon compression. The transition in the bonding character is also reflected in a significant change of the interatomic bond lengths. At a pressure of 30~GPa, the change of the FM to the NM state leads to a decrease in the Bi-Bi and Fe-Bi bonds from 2.95~\AA\, to 2.92~\AA\, and from 2.73~\AA\, to 2.69~\AA, respectively, while the Fe-Fe bond increases from 2.69~\AA\, to 2.77~\AA.

For isostructural compounds with lighter pnictogen elements Pn~=~\{P, As, Sb\}, the formation of Pn-dimers essentially leads to Zintl phases with semiconducting behavior~\cite{Armbruster2007}. In contrast, \ce{FeBi2} remains metallic although similar Bi dumbbells are formed. The Bi--Bi bond length of 2.92~{\AA} is slightly larger than the isolated double-bonded dianion \ce{[Bi\bond{=}Bi]^2-}~\cite{Xu2000}, which is about 2.84~{\AA}. This discrepancy can be attributed to extra electronic charge delocalized over the cations, in agreement with the antibonding states at the Fermi level of the dimers shown in Figure~\ref{fig:cohp}~(d). Therefore, the expected charge state is [Fe]$^{(2-\delta)+}$[\ce{Bi2}]$^{(2+\delta)-}$, where $\delta>0$. This non-integer charge can readily account for the metallic behavior of \ce{FeBi2} as opposed to the Zintl compounds where the octet rule implies a finite band gap as observed in \ce{FeAs2}~\cite{yannello_generality_2015}. Hence, despite the similarities in the main characteristics with other \ce{FePn2} compounds, metallic \ce{FeBi2} cannot be classified as a traditional Zintl phase.

\begin{figure}[tb]
\includegraphics[width=0.9\columnwidth]{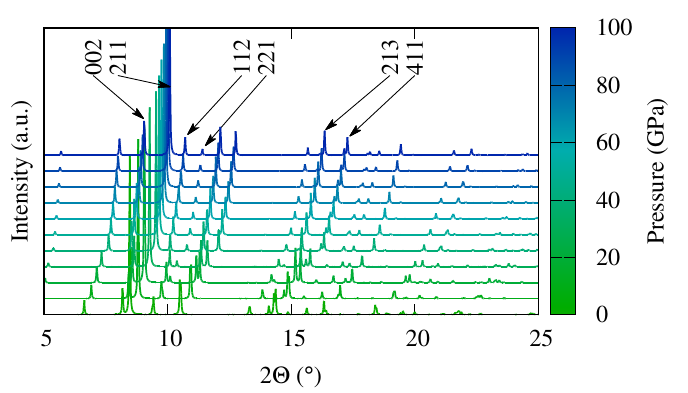}
\caption{\label{fig:MS_pxrd}  Simulated XRD  spectra of \ce{FeBi2} at  various pressures  for synchroton radiation at a wavelength of \SI{0.40663}{\angstrom}}
\end{figure}

Figure~\ref{fig:MS_pxrd} shows the evolution of the simulated X-ray diffraction (XRD) spectra as a function of pressure between 0 and 100~GPa. The change in bonding and the unit cell volume is reflected in the evolution of the XRD pattern, and the relative diffraction angles of the low index peaks could therefore serve as a fingerprint to indirectly distinguish the competing magnetic states. Specifically the pairs of reflections from $hkl=(002)/(211)$, $(112)/(221)$ and $(213)/(411)$ exhibit distinct changes in their relative positions around 30~GPa. In fact, preliminary XRD data has been recently collected with \textit{in-situ} high pressure synchrotron experiments in excellent agreement with our predictions, confirming the formation of the \ce{FeBi2} phase at high pressure. A detailed analysis of the experimental results will be published elsewhere.

\begin{figure}[tb]
\includegraphics[width=0.9\columnwidth]{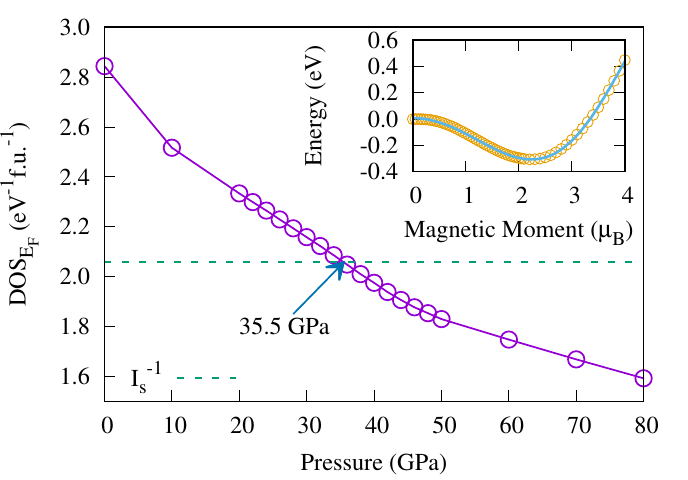}
\caption{The electronic DOS at the Fermi level DOS$_{E_F}$ as a function of pressure for the NM configuration. The dashed line denotes the value of the inverse Stoner parameter $I_s^{-1}$, and its intersection with the solid line indicates the transition pressure below which the magnetic state is preferred. The inset shows the total energy as a function of the magnetic moment $\mu_B$ at ambient pressure. The blue line indicates the fit to the polynomial function of order 6 which was used to extract the Stoner parameter $I_s$.}
\label{fig:stoner}
\end{figure}

\begin{figure}[tb]
\includegraphics[width=0.9\columnwidth]{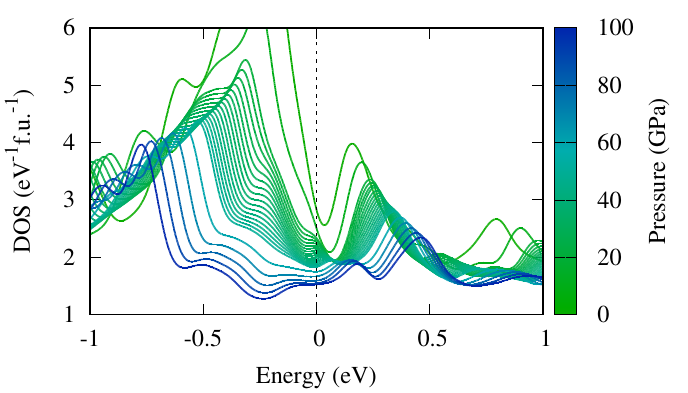}
\caption{The electronic DOS as a function of pressure in the non magnetic configuration, shifted such that the Fermi level is at zero. Note that the DOS at the Fermi level, DOS$_{E_F}$, gradually decreases as the pressure increases.}
\label{fig:dos}
\end{figure}

The magnetic collapse in FeBi$_{2}$ upon compression can be readily explained by the Stoner model~\cite{Stoner1938}, which is valid in the context of materials with itinerant magnetism~\cite{Cohen1997,Ortenzi2012,Jin2008,Sieberer2006}. According to this model, FM is favored if the gain in exchange energy is larger than the loss in kinetic energy~\cite{James1999}. The Stoner criterion serves as an indicator for this magnetic transition, which occurs if $\textrm{DOS}_{E_F}>I_{s}^{-1}$, where DOS$_{E_F}$ is the density of states at the Fermi level, and $I_{s}$ is the Stoner parameter which only weakly depends on the inter atomic distances~\cite{Zhang2016a}. The Stoner parameter can be obtained from a polynomial expansion of the energy as a function of the magnetic moment: $E(M)=E_{0} + a_{2}M^{2}+ a_{4}M^{4} \dots$, where $a_{2}=1/\textrm{DOS}_{E_F}^0\!-\!I_{s}$, and $\textrm{DOS}_{E_F}^0$ is the non magnetic $\textrm{DOS}_{E_F}$~\cite{kubler2000theory,James1999,hemley2002high} (see inset in Figure~\ref{fig:stoner}). With increasing pressure the value of DOS$_{E_F}$ gradually decreases (see Figure~\ref{fig:dos}), and at $p_c=35.5$~GPa the Stoner criterion is not satisfied anymore as shown in Figure~\ref{fig:stoner}, where $\textrm{DOS}_{E_F}<I_{s}^{-1}$, and thus the NM state is preferred for pressures above $p_c$. This result is in good agreement with the enthalpy plot shown in the top panel of Figure~\ref{fig:Magvol}, where NM becomes thermodynamically more favorable than FM above a pressure of 30~GPa, a value close to $p_c$.

\begin{table}[b]
\begin{ruledtabular}
  \centering
\begin{tabular} {c c c c}
Pressure (GPa) & $\lambda$  & $\omega_{\textrm{log}}$ (K) & $T_{\rm c}$ (K)\\
    \hline
40 & 0.50 & 184.6 & 1.3\\
60 & 0.41 & 217.9 & 0.5\\
80 & 0.35 & 244.3 & 0.1

\end{tabular} 
    \caption{    \label{tab:TC}Parameters derived from electron-phonon calculations at different pressures according to equation~\eqref{eq:mcmillan}.}
\end{ruledtabular}
\end{table}

\begin{figure}[tb]
\includegraphics[width=0.9\columnwidth]{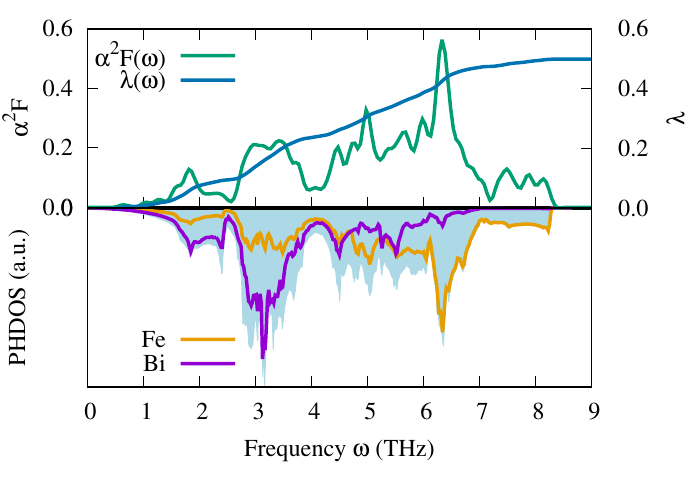}
\caption{\label{fig:ELPH} The electron-phonon coupling properties for FeBi$_2$ at 40~GPa. The Eliashberg spectral function $\alpha^2F(\omega)$ and the integrated coupling constant $\lambda(\omega)$ are shown in the top panel, whereas the partial PHDOS are shown in the lower panel, respectively. The shaded area indicates the total PHDOS.}
\end{figure}

Finally, we estimate the superconducting temperature of FeBi$_2$ in its NM state at 40~GPa within the Bardeen-Cooper-Schrieffer (BCS) theory. The Eliashberg spectral function, the coupling constant $\lambda$ and the phonon density of states (PHDOS) are shown in Figure~\ref{fig:ELPH}. According to our calculations, FeBi$_2$ is a superconductor with $T_{\rm c}=1.3$~K and a moderate electron-phonon coupling constant of $\lambda=0.50$. The lower panel in Figure~\ref{fig:ELPH} shows the total PHDOS together with the partial, atom projected PHDOS. By comparing the spectral function $\alpha^2F(\omega)$ and the frequency dependent coupling constant $\lambda(\omega)$ with the partial PHDOS we conclude that there are two major contributions to the final value of $\lambda$. First, there is a strong increase in $\lambda(\omega)$ at a frequency of around $\omega=3$~THz, which arises mainly from the Bi vibrations. Second, there is an additional strong contribution to $\lambda(\omega)$ in a frequency range between $\omega=4.5-7$~THz, which can be attributed to the Fe dominated region of the PHDOS. Table~\ref{tab:TC} contains the results of the electron-phonon coupling calculations at two additional pressures, 60 and 80~GPa. The electron-phonon coupling strength decreases with increasing pressure, leading to a supression of the superconducting transition temperatures, a behavior also observed in other bismuth superconductors (e.g. \ce{CaBi3}~\cite{dong_rich_2015}). This trend in $T_{\rm c}$ can be readily explained by the decreasing DOS$_{E_F}$ shown in Figure~\ref{fig:stoner}, since mainly electrons at the Fermi surface contribute to the electron-phonon coupling.  

Although it is in principle possible for any metal to attain superconductivity at low temperatures, superconducting behavior is usually suppressed in ferromagnetic materials and only few examples have been reported where superconductivity coexists with intrinsic magnetism~\cite{saxena_superconductivity_2000,aoki_coexistence_2001,huy_superconductivity_2007}. Elemental, non magnetic hcp-iron shows superconductivity above 13~GPa with a maximum $T_{\rm c}=2$~K at 20~GPa~\cite{shimizu_superconductivity_2001}, and superconductivity in other iron containing materials at the border of magnetism such as FeSe~\cite{mizuguchi_superconductivity_2008,hsu_superconductivity_2008} cannot be fully explained by conventional BCS theory, where the conventional $T_{\rm c}$ is about one order of magnitude lower than the experimental values~\cite{subedi_density_2008,boeri_is_2008,mazin_unconventional_2008}. Similarly, LiFeAs was found to superconduct at 18\,K, while the $T_{\rm c}$ from BCS theory is less than 1\,K~\cite{Jishi2010}. Since electron-phonon coupling cannot fully account for the observed superconducting behavior in above materials, spin fluctuation has been considered as a possible coupling mechanism~\cite{jarlborg_ferromagnetic_2002,Mazin2010,mazin_unconventional_2008}. Therefore, since \ce{FeBi2} is at the verge of FM and AFM order it could possibly exhibit unconventional superconductivity, in which case the computed $T_{\rm c}$ is merely a probable lower limit of the real value.

\section{\label{sec:con}Conclusion}

In summary, we have successfully predicted the stability and superconducting properties of the first binary compound in the ambient-immiscible Fe--Bi system at high pressure, \ce{FeBi2}. It crystallizes in the \ce{Al2Cu} structure with space group \textit{I}4/\textit{mcm}, is thermodynamically stable above 37.5~GPa and undergoes a series of magnetic transitions upon compression: from ferromagnetic ordering at ambient pressure to an anti-ferromagnetic state and finally to a non magnetic configuration at pressures above 38~GPa. These magnetic transitions are accompanied by structural changes, where short, covalent Bi-Bi bonds are formed in the non magnetic state at high pressure, leading to a significant decrease in the unit cell volume. The resulting low $pV$ term in the enthalpy is thus the main driving force responsible for the formation of \ce{FeBi2}. Electron-phonon coupling calculations show that \ce{FeBi2} is a potential superconductor with a moderate coupling constant and a critical temperature of $T_{\rm c}=1.3$~K at 40~GPa. However, the magnetic frustration in \ce{FeBi2} might be an indication of non-conventional superconductivity with a higher value of $T_{\rm c}$.

\section*{Acknowledgments}

We thank J.A. Flores-Livas and V. Hegde for valuable discussions. M.A. acknowledges support from the Novartis Universit{\"a}t Basel Excellence Scholarship for Life Sciences and the Swiss National Science Foundation (P300P2-158407). S.~S.~N. and C.W. acknowledge support by the U.S. Department of Energy, Office of Science, Basic Energy Sciences, under Grant DE-FG02-07ER46433. The Swiss National Supercomputing Center in Lugano (Project s499 and s621), the Extreme Science and Engineering Discovery Environment (XSEDE) (which is supported by National Science Foundation grant number OCI-1053575), the Bridges system at the Pittsburgh Supercomputing Center (PSC) (which is supported by NSF award number ACI-1445606), the Quest high performance computing facility at Northwestern University, and the National Energy Research Scientific Computing Center (DOE: DE-AC02-05CH11231), are gratefully acknowledged.

%\bibliography{febi}

\end{document}